\documentstyle[fleqn,iopfts,12pt]{ioplppt}
\jl{1}
\eqnobysec
\begin{document}
\title{Coherent states for a particle on a sphere}
\author{K Kowalski and  J Rembieli\'nski}
\address{Department of Theoretical Physics, University
of \L\'od\'z, ul.\ Pomorska 149/153,\\ 90-236 \L\'od\'z,
Poland}
\begin{abstract}
The coherent states for a particle on a sphere are introduced.
These states are labelled by points of the classical phase space, that is
the position on the sphere and the angular momentum of a particle.
As with the coherent states for a particle on a circle discussed in
Kowalski K {\em et al\/} 1996 {\em J. Phys. A} {\bf 29} 4149, we deal
with a deformation of the classical phase space related with quantum
fluctuations.  The expectation values of the position and the
angular momentum in the coherent states are regarded as the best
possible approximation of the classical phase space.  The correctness
of the introduced coherent states is illustrated by an example of
the rotator.
\end{abstract}
\pacs{02.20.Sv, 02.30.Gp, 02.40.-k, 03.65.-w, 03.65.Sq}
\section{Introduction}
It has become a clich\'e to say that coherent states abound in
quantum physics \cite{1}.  Moreover, it turns out that they can also
be applied in the theory of quantum deformations \cite{2} and even
in the theory of classical dynamical systems \cite{3}.

In spite of the fact that the problem of the quantization of a
particle motion on a sphere is at least seventy years old, there still
remains an open question concerning 
the coherent states for a particle on a sphere.  Indeed, the
celebrated spin coherent states introduced by Radcliffe \cite{4} and
Perelomov \cite{5} are labelled by points of a sphere, i.e., the
elements of the configuration space.  On the other hand, it seems
that as with the standard coherent states, the coherent states for a
particle on a sphere should be marked with points of the phase
space rather than the configuration space.

The aim of this work is to introduce the coherent states for a quantum
particle on the sphere $S^2$, labelled by points of
the phase space, that is the cotangent bundle $T^*S^2$.  The construction
follows the general scheme introduced in \cite{6} for the case of the motion
in a circle, based on the polar decomposition of the operator
defining via the eigenvalue equation the coherent states.   From the technical
point of view our treatment utilizes both the
Barut-Girardello \cite{7} and Perelomov approach \cite{5}.  Namely, as with
the Barut-Girardello approach the coherent states are defined as the
eigenvectors of some non-Hermitian operators.  On the other hand, in
analogy to the Perelomov formalism those states are generated from
some ``vacuum vector'', nevertheless in opposition to the
Perelomov group-theoretic construction, the coherent states are
obtained by means of the non-unitary action.

In section 2 we recall the construction of the coherent states for
a particle on a circle.  Sections 3--6 are devoted to the
definition of the coherent states for a particle on a sphere
and discussion of their most important properties.  For an easy
illustration of the introduced approach we study in section 7 the
case with the free motion on a sphere.
\section{Coherent states for a particle on a circle}
In this section we recall the basic properties of the coherent
states for a particle on a circle introduced in \cite{6}.
Consider the case of the free motion in a circle.  For the sake of
simplicity we assume that the particle has unit mass and it moves in
a unit circle.  The classical Lagrangian is
\begin{equation}
L = \hbox{$\scriptstyle1\over2$}\dot \varphi^2,
\end{equation}
so the angular momentum canonically conjugate to the angle $\varphi$
is given by
\begin{equation}
J = \frac{\partial L}{\partial \dot \varphi}=\dot \varphi,
\end{equation}
and the Hamiltonian can be written as
\begin{equation}
H = \hbox{$\scriptstyle1\over2$}J^2.
\end{equation}
Evidently, we have the Poissson bracket of the form
\begin{equation}
\{\varphi,J\} = 1,
\end{equation}
implying accordingly to the rules of the canonical
quantization the commutator
\begin{equation}
[\hat\varphi,\hat J] = i,
\end{equation}
where we set $\hbar=1$. The operator $\hat\varphi$ does not take
into account the topology of the circle and (2.5) needs very subtle
analysis.  The better candidate to represent the position of the
quantum particle on the unit circle is the unitary operator $U$
\begin{equation}
U = e^{i\hat\varphi}.
\end{equation}
Indeed, the substitution $\hat\varphi\to\hat\varphi+2n\pi$ does not
change $U$, i.e.\ $U$ preserves the topology of the circle.  The operator
$U$ leads to the algebra
\begin{equation}
[\hat J,U] = U,
\end{equation}
where $U$ is unitary.  Consider the eigenvalue equation
\begin{equation}
\hat J|j\rangle = j|j\rangle.
\end{equation}
Using (2.7) and (2.8) we find that the operators $U$ and
$U^\dagger $ are the ladder operators, namely
\numparts
\begin{eqnarray}
U|j\rangle &=& |j+1\rangle,\\
U^\dagger |j\rangle &=& |j-1\rangle.
\end{eqnarray}
\endnumparts
Demanding the time-reversal invariance of representations of the
algebra (2.7) we conclude \cite{6} that the eigenvalues $j$ of the operator
$\hat J$ can be only integer (boson case) or half-integer (fermion case).

We define the coherent states $|\xi\rangle$ for a particle on a circle
by means of the eigenvalue equation
\begin{equation}
Z|\xi\rangle = \xi|\xi\rangle,
\end{equation}
where $\xi$ is complex. In analogy to the eigenvalue equation
satisfied by the standard coherent states $|z\rangle$ \cite{8,9} with complex
$z$, of the form
\begin{equation}
e^{i\hat a}|z\rangle = e^{iz}|z\rangle,
\end{equation}
where $\hat a\sim \hat q+i\hat p$ is the standard Bose
annihilation operator and $\hat q$ and $\hat p$ are the
position and momentum operators, respectively, we set
\begin{equation}
Z := e^{i(\hat \varphi + i\hat J)}.
\end{equation}
Hence, making use of the Baker-Hausdorff formula we get
\begin{equation}
Z = e^{-\hat J + \hbox{$\frac{1}{2}$}}U.
\end{equation}
We remark that the complex number $\xi$ should parametrize the
cylinder which is the classical phase space for the particle
moving in a circle.  The convenient parametrization of $\xi$
consistent with the form of the operator $Z$ such that
\begin{equation}
\xi = e^{-l + i\varphi}.
\end{equation}
arises from the deformation of the circular cylinder by means of the
transformation
\begin{equation}
x=e^{-l}\cos\varphi,\qquad y=e^{-l}\sin\varphi,\qquad z=l.
\end{equation}
The coherent states $|\xi\rangle$ can be represented as
\begin{equation}
|\xi\rangle = e^{-(\ln\xi) \hat J}|1\rangle,
\end{equation}
where
\begin{equation}
|1\rangle =
\sum_{j=-\infty}^{\infty}e^{-\frac{j^2}{2}}|j\rangle.
\end{equation}
The coherent states satisfy
\begin{equation}
\frac{\langle \xi|\hat J|\xi\rangle}{\langle
\xi|\xi\rangle}\approx l,
\end{equation}
where the maximal error arising in the case $l\to0$ is of order
$0.1$ per cent and we have the exact equality in the case with $l$
integer or half-integer.  Therefore, $l$ can be identified with the
classical angular momentum.  Furthermore, we have
\begin{equation}
\frac{\langle \xi|U|\xi\rangle}{\langle \xi|\xi\rangle}
 \approx
e^{-\frac{1}{4}}e^{i\varphi}.
\end{equation}
It thus appears that the average value of $U$ in the normalized
coherent state does not belong to the unit circle.  On introducing
the relative average of $U$ of the form
\begin{equation}
\frac{\langle U\rangle_{\xi}}{\langle U\rangle_{\eta}} :=
\frac{\langle \xi|U|\xi\rangle}{\langle \eta|U|\eta\rangle},
\end{equation}
where $|\xi\rangle$ and $|\eta\rangle$ are the normalized
coherent states, we find
\begin{equation}
\frac{\langle U\rangle_{\xi}}{\langle U\rangle_1}
\approx e^{i\varphi}.
\end{equation}
From (2.21) it follows that that the relative expectation
value $\langle U\rangle_{\xi}/\langle U\rangle_1$ is the
most natural candidate to describe the average position of a
particle on a circle and $\varphi$ can be regarded as the
classical angle.

We remark that the coherent states on the circle have been
recently discussed by Gonz\'ales {\em et al\/} \cite{10}.  In spite
of the fact that they formally generalize the coherent states
described above, the ambiguity of the definition of those states
manifesting in their dependence on some extra parameter, can be
avoided only by demanding the time-reversal invariance mentioned
earlier, which leads precisely to the coherent states introduced in
\cite{6}.  Since the time-reversal symmetry seems to be fundamental
one for the motion of the classical particle in a circle and makes
the quantization unique, therefore the generalization of the coherent
states discussed in \cite{10} which does not preserve that symmetry is
of interest rather from the mathematical point of view.

Having in mind the properties of the standard coherent states one
may ask about the minimalization of the Heisenberg uncertainty
relations by the introduced coherent states for a particle
on a circle.  In our opinion, in the case with the compact manifolds
the minimalization of the Heisenberg uncertainty relations is not an
adequate tool for the definition of the coherent states.  A
counterexample can be easily deduced from (2.7), (2.8) and (2.9).
Indeed, taking into account (2.8) and (2.9) we find that for the
eigenvectors $|j\rangle$'s of the angular momentum $\hat J$ the
equality sign is attended in the Heisenberg uncertainty relations
implied by (2.7) such that
\begin{equation}
(\Delta \hat J)^2\ge\frac{1}{4}\frac{|\langle U \rangle|^2}{1-|\langle U
\rangle|^2}.
\end{equation}
More precisely, for these states (2.22) takes the form $0=0$.
On the other hand, the vectors $|j\rangle$'s are clearly rather poor
candidate for the coherent states.  In our opinion the fact that the
coherent states are ``the most classical'' ones is better described
by the following easily proven formulae:
\begin{eqnarray}
(\Delta \hat J)^2&\approx& {\rm const},\\
\frac{\langle U^2\rangle}{\langle U\rangle^2}&\approx&{\rm const},
\end{eqnarray}
where the approximations are very good ones.  In fact, these
relations mean that the quantum variables $\hat J$ and $U$ are at
practically constant ``distance'' from their classical counterparts
$\langle \hat J\rangle$ and $\langle U\rangle$, respectively, and
therefore the quantum observables and the corresponding expectation
values connected to the classical phase space are mutually related.
We point out that in the case with the standard coherent states for
a particle on a real line we have the exact formulae
\begin{eqnarray}
(\Delta \hat p)^2&=&{\rm const},\\
(\Delta \hat q)^2&=&{\rm const}.
\end{eqnarray}
It seems to us that the approximative nature of the relations (2.23)
and (2.24) is related to the compactness of the circle.
\section{Unitary representations of the $e(3)$ algebra and quantum
mechanics on a sphere}
Our experience with the case of the circle discussed in the previous
section indicates that in order to introduce the coherent states we
should first identify the algebra adequate for the study of the
motion on a sphere.  The fact that the algebra (2.7) referring to the
case with the circle $S^1$ is equivalent to the $e(2)$ algebra, where
$E(2)$ is the group of the plane consisting of translations and rotations,
\begin{equation}
[\hat J,X_\alpha]={\rm i}\varepsilon_{\alpha\beta}X_\beta,
\qquad [X_\alpha,X_\beta]=0,\qquad \alpha,\,\beta=1,\,2,
\end{equation}
realized in a unitary irreducible representation by Hermitian operators
\begin{equation}
X_1=r(U+U^\dagger)/2,\qquad X_2=r(U-U^\dagger)/2{\rm i},
\end{equation}
where the Casimir is
\begin{equation}
X_1^2+X_2^2=r^2,
\end{equation}
and $\varepsilon_{\alpha\beta}$ is the anti-symmetric tensor,
indicates that the most natural algebra for the case with the
sphere $S^2$  is the $e(3)$ algebra such that
\begin{equation}
[J_i,J_j]={\rm i}\varepsilon_{ijk}J_k,\qquad [J_i,X_j]={\rm i}
\varepsilon_{ijk}X_k,\qquad [X_i,X_j]=0,\qquad i,\,j,\,k=1,\,2,\,3.
\end{equation}
Indeed, the algebra (3.4) has two Casimir operators given in a unitary
irreducible representation by
\begin{equation}
{\bi X}^2=r^2,\qquad {\bi J}\bdot{\bi X}=\lambda,
\end{equation}
where dot designates the scalar product.  Therefore, as with the generators
$X_\alpha $, $\alpha=1,\,2$, describing the position of a particle
on the circle, the generators $X_i$, $i=1,\,2,\,3$, can be regarded as
quantum counterparts of the Cartesian coordinates of the points of the sphere
$S^2$ with radius $r$.  We point out that unitary irreducible
representations of (3.4) can be labelled by $r$ and the new scale
invariant parameter $\zeta =\frac{\lambda }{r}$.  It is clear that $\zeta $
is simply the projection of the angular momentum ${\bi J}$ on the
direction of the radius vector of a particle.  Since we did not find
any denomination for such an entity in the literature, therefore we have
decided to call $\zeta $ the {\em twist\/} of a particle.

Let us now recall the basic properties of the unitary
representations of the $e(3)$ algebra.  The $e(3)$ algebra expressed with
the help of operators $J_3$, $J_\pm=J_1\pm {\rm i}J_2$, $X_3$ and $X_\pm=X_1\pm
{\rm i}X_2$, takes the form
\numparts
\begin{eqnarray}
[J_+,J_-] &=& 2J_3,\qquad [J_3,J_\pm]=\pm J_\pm,\\
{}[J_\pm ,X_\mp] &=& \pm 2X_3,\qquad [J_\pm,X_\pm]=0,\qquad [J_\pm ,X_3]=\mp X_\pm,\\
{}[J_3,X_\pm] &=& \pm X_\pm,\qquad [J_3,X_3]=0,\\
{}[X_+,X_-] &=& [X_\pm,X_3]=0.
\end{eqnarray}
\endnumparts
Consider the irreducible representation of the above algebra in the
angular momentum basis spanned by the common eigenvectors
$|j,m;r,\zeta\rangle$ of the operators ${\bi J}^2=J_+J_-+J_3^2-J_3$,
$J_3$, ${\bi X}^2$ and ${\bi J}\bdot{\bi X}/r$
\numparts
\begin{eqnarray}
&&{\bi J}^2 |j,m;r,\zeta\rangle = j(j+1) |j,m;r,\zeta\rangle,\qquad J_3
|j,m;r,\zeta\rangle=m|j,m;r,\zeta\rangle,\\
&&{\bi X}^2 |j,m;r,\zeta\rangle=r^2 |j,m;r,\zeta\rangle,\qquad
({\bi J}\bdot{\bi X}/r) |j,m;r,\zeta\rangle=\zeta|j,m;r,\zeta\rangle,
\end{eqnarray}
\endnumparts
where $-j\le m\le j$.  Recall that
\begin{equation}
J_\pm |j,m;r,\zeta\rangle=\sqrt{(j\mp m)(j\pm m+1)}\,|j,m\pm 1;r,\zeta\rangle.
\end{equation}
The operators $X_\pm$ and $X_3$ act on the vectors $|j,m;r,\zeta\rangle$ in the
following way:
\numparts
\begin{eqnarray}
X_+ |j,m;r,\zeta\rangle
&=&-\frac{r\sqrt{(j+1)^2-\zeta^2}\sqrt{(j+m+1)(j+m+2)}}
{(j+1)\sqrt{(2j+1)(2j+3)}}|j+1,m+1;r,\zeta\rangle\nonumber\\
&&{}+\frac{\zeta r\sqrt{(j-m)(j+m+1)}}{j(j+1)}|j,m+1;r,\zeta\rangle\nonumber\\
&&{}+\frac{r\sqrt{j^2-\zeta^2}\sqrt{(j-m-1)(j-m)}}{j\sqrt{(2j-1)(2j+1)}}
|j-1,m+1;r,\zeta\rangle,\\
X_- |j,m;r,\zeta\rangle
&=&\frac{r\sqrt{(j+1)^2-\zeta^2}\sqrt{(j-m+1)(j-m+2)}}
{(j+1)\sqrt{(2j+1)(2j+3)}}|j+1,m-1;r,\zeta\rangle\nonumber\\
&&{}+\frac{\zeta r\sqrt{(j-m+1)(j+m)}}{j(j+1)}|j,m-1;r,\zeta\rangle\nonumber\\
&&{}-\frac{r\sqrt{j^2-\zeta^2}\sqrt{(j+m-1)(j+m)}}{j\sqrt{(2j-1)(2j+1)}}
|j-1,m-1;r,\zeta\rangle,\\
X_3 |j,m;r,\zeta\rangle
&=&\frac{r\sqrt{(j+1)^2-\zeta^2}\sqrt{(j-m+1)(j+m+1)}}
{(j+1)\sqrt{(2j+1)(2j+3)}}|j+1,m;r,\zeta\rangle\nonumber\\
&&\fl\fl{}+\frac{\zeta rm}{j(j+1)}|j,m;r,\zeta\rangle+
\frac{r\sqrt{j^2-\zeta^2}\sqrt{(j-m)(j+m)}}{j\sqrt{(2j-1)(2j+1)}}|j-1,m;r,\zeta
\rangle.
\end{eqnarray}
\endnumparts
An immediate consequence of (3.9) is the existence of the minimal
$j=j_{\rm min}$ satisfying
\begin{equation}
j_{\rm min}=|\zeta| .
\end{equation}
Thus, it turns out that in the representation defined by (3.9) the twist
$\zeta $ can be only integer or half integer.  We finally write down
the orthogonality and completeness conditions satisfied by the
vectors $|j,m;r,\zeta\rangle$ such that
\begin{eqnarray}
&&\langle j,m;r,\zeta|j',m';r,\zeta\rangle=\delta_{jj'}\delta_{mm'},\\
&&\sum_{j=|\zeta|}^{\infty}\sum_{m=-j}^{j}
|j,m;r,\zeta\rangle\langle j,m;r,\zeta|=I,
\end{eqnarray}
where $I$ is the identity operator.
\section{Definition of coherent states for a particle on a sphere}
Now, an experience with the circle indicates that one should identify by means
of the $e(3)$ algebra an analogue of the unitary operator $U$ (2.6),
representing the position of a particle on a sphere.  To do
this, let us recall that a counterpart of the ``position'' $e^{{\rm
i}\varphi}$ on the circle $S^1$ is a unit length imaginary quaternion
which can be represented with the help of the Pauli matrices
$\sigma_i$, $i=1,\,2,\,3$, as
\begin{equation}
\eta = {\rm i}{\bi n}\bdot{\bsigma},
\end{equation}
where ${\bi n}^2=1$.  Notice that $\eta$ is simply an element of the
$SU(2)$ group and it is related to the $S^2\approx SU(2)/U(1)$
quotient space.  Therefore the most natural choice for the
``position operator'' of a particle on a sphere is to set
\begin{equation}
V=\hbox{$\scriptstyle 1\over r$}\bsigma\bdot{\bi X},
\end{equation}
where $X_i$, $i=1,\,2,\,3$
obey (3.4) and (3.9) and we have omitted for convenience the imaginary
factor i.  Furthermore, let us introduce a version of the Dirac matrix
operator \cite{11}
\begin{equation}
K := -(\bsigma\bdot{\bi J}+1).
\end{equation}
Observe that
\begin{equation}
V^\dagger=V,\qquad K^\dagger=K.
\end{equation}
Making use of the operators $V$ and $K$ we can write the relations
defining the $e(3)$ algebra in the space of the unitary irreducible
representation introduced above as
\numparts
\begin{eqnarray}
({\rm Tr}\bsigma K)^2 &=& 4K(K+1),\\
{}[K,V]_+ &=& {\rm Tr}KV,\\
V^2&=&I,
\end{eqnarray}
\endnumparts
where ${\rm Tr}A=A_{11}+A_{22}$, and the subscript ``+'' designates the
anti-commutator.  In particular,
\begin{equation}
{\rm Tr}KV=-2{\bi J}\bdot{\bi X}/r=-2\zeta .
\end{equation}
It should also be noted that in view of (4.4) and (4.5{\em c}) $V$
satisfies the unitarity condition $V^\dagger V=I$.

We now introduce the vector operator ${\bi Z}$ generating, via the eigenvalue
equation analogous to (2.10), the  coherent states for a particle on
a sphere $S^2$.  The experience with the circle (see eq.\ (2.13)) suggests the
following form of the ``polar decomposition'' for the matrix operator
counterpart $Z$ of the operator ${\bi Z}$:
\begin{equation}
Z=e^{-K}V.
\end{equation}
Indeed, it is easy to see that in the case of the circular motion in
the equator defined semiclassically by $J_1=J_2=0$ and $X_3=0$, $Z$
reduces to the diagonal matrix operator with $Z$ given by (2.13) and its
Hermitian conjugate on the diagonal.  Furthermore, using (4.5{\em b})
we find
\begin{equation}
Z-Z^{-1} = 2\zeta K^{-1}\sinh K.
\end{equation}
Motivated by the complexity of the problem we now restrict to the
simplest case of the twist $\zeta=0$ when (4.8) takes the form
\begin{equation}
Z^2=I.
\end{equation}
In the following we confine ourselves to the case $\zeta=0$.
The general case with arbitrary $\zeta\ne0$ will be discussed in a
separate work.  Besides (4.9) we have also remarkably simple
relation (4.5{\em b}) referring to $\zeta=0$ such that
\begin{equation}
[K,V]_+ = 0.
\end{equation}
Notice that the
case $\zeta=0$ is the ``most classical'' one.  Indeed, the projection
of the angular momentum onto the direction of the radius vector should
vanish for the classical particle on a sphere.  It should also be noted
that in view of (3.10) $j$'s and $m$'s labelling the basis vectors
$|j,m;r,\zeta\rangle$ are integer in the case of the twist $\zeta =0$.
We finally point out that the condition $\zeta=0$ ensures the
invariance of the irreducible representation of the $e(3)$ algebra
under time inversions and parity transformations which
change the sign of the product ${\bi J}\bdot{\bi X}$.  Clearly
demanding the time-reversal or the parity invariance when $\zeta\ne0$ one
should work with representations involving both $\zeta$ and $-\zeta$.

We now return to (4.7).  Making use of (4.10) and the fact that the
matrix operator $V$ in view of (4.2) is traceless one we obtain for $\zeta=0$
\begin{equation}
{\rm Tr}Z=0.
\end{equation}
Hence,
\begin{equation}
Z = \bsigma\bdot{\bi Z}.
\end{equation}
Taking into account (4.9) we get from (4.12)
\begin{equation}
{\bi Z}^2=1,
\end{equation}
and
\begin{equation}
[Z_i,Z_j]=0,\qquad i,j=1,\,2,\,3.
\end{equation}
As with (4.2) describing in the matrix language the position of a quantum
particle on a sphere, the matrix operator (4.12) can be only interpreted as a
convenient arrangement of the operators $Z_i$ generating the coherent
states, simplifying the algebraic analysis of the problem.  Accordingly, we
define the coherent states for a quantum mechanics on a sphere in terms of
operators $Z_i$, as the solutions of the eigenvalue equation such that
\begin{equation}
{\bi Z} |{\bi z}\rangle = {\bi z} |{\bi z}\rangle,
\end{equation}
where in view of (4.13) ${\bi z}^2=1$.  What is ${\bi Z}$ ?  Using
(4.7), (4.2), (4.3) and setting $\zeta=0$, we find after some calculation
\begin{eqnarray}
{\bi Z} &=&\left(\frac{e^{\frac{1}{2}}}{\sqrt{1+4{\bi J}^2}}{\rm
sinh}\hbox{$\scriptstyle 1\over2 $}\sqrt{1+4{\bi
J}^2}+e^{\frac{1}{2}}{\rm cosh}\hbox{$\scriptstyle 1\over2 $}
\sqrt{1+4{\bi J}^2}\right){{\bi X}\over r}\nonumber\\
&&{}+{\rm i}\left(\frac{2e^{\frac{1}{2}}}{\sqrt{1+4{\bi J}^2}}{\rm sinh}
\hbox{$\scriptstyle 1\over2 $}\sqrt{1+4{\bi J}^2}\right){\bi
J}\times{{\bi X}\over r}.
\end{eqnarray}
We remark that $Z_i$ have the structure resembling
the standard annihilation operators.  In fact, one can easily check
that it can be written as a combination
\begin{equation}
{\bi Z}=a{\bi X}+{\rm i}b{\bi P},
\end{equation}
of the ``position operator'' ${\bi X}$ and the ``momentum'' ${\bi P}$, where
the coefficients $a$ and $b$ are functions of ${\bi J}^2$.  We finally point
out that derivation of the operator ${\bi Z}$ (4.16) without the knowledge of
the matrix operator $Z$ seems to be very difficult task.
\section{Construction of the coherent states}
In this section we construct the coherent states specified by the eigenvalue
equation (4.15).  On projecting (4.15) on the basis vectors
$|j,m;r\rangle\equiv|j,m;r,0\rangle$ and using (3.7{\em a}), (3.8) and (3.9) with
$\zeta=0$ we arrive at the system of linear difference equations satisfied by
the Fourier coefficients of the expansion of the coherent state
$|{\bi z}\rangle$ in the basis $|j,m;r\rangle$.  The direct solution of such
system in the general case seems to be difficult task.  Therefore, we adopt the
following technique.  We first solve the eigenvalue equation for ${\bi z}=
{\bi n}_3=(0,0,1)$, and then generate the coherent states from the vector
${\bi n}_3$ using the fact (see (4.16)) that ${\bi Z}$ is a vector operator.  As
demonstrated in the next section the case with ${\bi z}={\bi n}_3$ refers to
${\bi x}=(0,0,1)$ and ${\bi l}={\bf 0}$, where ${\bi x}$ is the
position and ${\bi l}$ the angular momentum, respectively, i.e., the particle
resting on the ``North Pole'' of the sphere.  Let us write down the
eigenvalue equation (4.15) for ${\bi z}={\bi n}_3$
\begin{equation}
{\bi Z} |{\bi n}_3\rangle={\bi n}_3 |{\bi n}_3\rangle.
\end{equation}
Using the following relations which can be easily derived with the help
of (4.16), (3.7{\em a}), (3.8) and (3.9) with $\zeta =0$:
\numparts
\begin{eqnarray}
Z_1 |j,m;r\rangle
&=&-\frac{1}{2}e^{-j-1}\sqrt{\frac{(j+m+1)(j+m+2)}{(2j+1)(2j+3)}}
|j+1,m+1;r\rangle\nonumber\\
&&{}+\frac{1}{2}e^j\sqrt{\frac{(j-m-1)(j-m)}{(2j-1)(2j+1)}}
|j-1,m+1;r\rangle\nonumber\\
&&+\frac{1}{2}e^{-j-1}\sqrt{\frac{(j-m+1)(j-m+2)}{(2j+1)(2j+3)}}
|j+1,m-1;r\rangle\nonumber\\
&&{}-\frac{1}{2}e^j\sqrt{\frac{(j+m-1)(j+m)}{(2j-1)(2j+1)}}
|j-1,m-1;r\rangle,\\
Z_2 |j,m;r\rangle
&=&\frac{{\rm i}}{2}e^{-j-1}\sqrt{\frac{(j+m+1)(j+m+2)}{(2j+1)(2j+3)}}
|j+1,m+1;r\rangle\nonumber\\
&&{}-\frac{{\rm i}}{2}e^j\sqrt{\frac{(j-m-1)(j-m)}{(2j-1)(2j+1)}}
|j-1,m+1;r\rangle\nonumber\\
&&+\frac{{\rm i}}{2}e^{-j-1}\sqrt{\frac{(j-m+1)(j-m+2)}{(2j+1)(2j+3)}}
|j+1,m-1;r\rangle\nonumber\\
&&{}-\frac{{\rm i}}{2}e^j\sqrt{\frac{(j+m-1)(j+m)}{(2j-1)(2j+1)}}
|j-1,m-1;r\rangle,\\
Z_3 |j,m;r\rangle
&=&e^{-j-1}\sqrt{\frac{(j-m+1)(j+m+1)}{(2j+1)(2j+3)}}
|j+1,m;r\rangle\nonumber\\
&&{}+e^j\sqrt{\frac{(j-m)(j+m)}{(2j-1)(2j+1)}}|j-1,m;r\rangle,
\end{eqnarray}
\endnumparts
it can be easily checked that the solution to (5.1) is given by
\begin{equation}
|{\bi
n}_3\rangle=\sum_{j=0}^{\infty}e^{-\frac{1}{2}j(j+1)}\sqrt{2j+1}|j,0;r\rangle.
\end{equation}
Now, using the commutator
\begin{equation}
[{\bi w}\bdot{\bi J},{\bi Z}]=-{\rm i}{\bi w}\times{\bi Z},
\end{equation}
where ${\bi w}\in{\Bbb C}^3$, we generate the complex rotation of
${\bi Z}$
\begin{equation}
e^{{\bi w}\bdot{\bi J}}{\bi Z}e^{-{\bi w}\bdot{\bi J}}=
\cosh\sqrt{{\bi w}^2}\,{\bi Z}-{\rm i}\frac{\sinh\sqrt{{\bi w}^2}}
{\sqrt{{\bi w}^2}}
{\bi w}\times{\bi Z}+\frac{1-\cosh\sqrt{{\bi w}^2}}{{\bi w}^2}{\bi w}
({\bi w}\bdot{\bi Z}).
\end{equation}
Taking into account (5.5) and (4.15) we find that the coherent states
can be expressed by
\begin{equation}
|{\bi z}\rangle = e^{{\bi w}\bdot{\bi J}}|{\bi n}_3\rangle,
\end{equation}
where ${\bi w}$ is given by
\begin{equation}
{\bi w}=\frac{{\rm arccosh}z_3}{\sqrt{1-z_3^2}}{\bi z}\times{\bi n}_3.
\end{equation}
It thus appears that the coherent states can be written as
\begin{equation}
|{\bi z}\rangle = \exp\left[\frac{{\rm arccosh}z_3}{\sqrt{1-z_3^2}}
({\bi z}\times{\bi n}_3)\bdot{\bi J}\right]
|{\bi n}_3\rangle.
\end{equation}
We remark that the discussed coherent states are generated
analogously as in the case of the circle described by the equation
(2.16).  The formula (5.8) can be furthermore written in the form
\begin{equation}
|{\bi z}\rangle = e^{\mu J_-}e^{\gamma J_3}e^{\nu J_+} |{\bi
n}_3\rangle,
\end{equation}
where
\begin{equation}
\mu =\frac{z_1+{\rm i}z_2}{1+z_3},\qquad \nu=\frac{-z_1+{\rm
i}z_2}{1+z_3},\qquad \gamma =\ln\frac{1+z_3}{2}.
\end{equation}
Finally, eqs.\ (5.9), (5.3), (3.7{\em a}) and (3.8) taken together yield
the following formula on the coherent states:
\begin{equation}
\fl |{\bi z}\rangle =\sum_{j=0}^{\infty}e^{-\frac{1}{2}j(j+1)}
\sqrt{2j+1}\sum_{m=0}^{j}\frac{\nu^m}{m!}\frac{(j+m)!}{(j-m)!}
e^{\gamma m}\sum_{k=0}^{j+m}\frac{\mu^k}{k!}
\sqrt{\frac{(j-m+k)!}{(j+m-k)!}} |j,m-k;r\rangle,
\end{equation}
where $\mu ,\,\nu$ and $\gamma$ are expressed by (5.10) and ${\bi
z}^2=1$.  Taking into account the identities
\begin{equation}
\sum_{s=0}\sp{n}\frac{(s+k)!}{(s+m)!s!(n-s)!}z^s=\frac{k!}{m!n!}\,\,
{}_2F_1(-n,k+1,m+1;-z),
\end{equation}
\begin{equation}
C_n^\alpha(x) =\frac{\Gamma(n+2\alpha)}{\Gamma(n+1)\Gamma(2\alpha)}
\,{}_2F_1(-n,n+2\alpha,\alpha+\hbox{$\scriptstyle 1\over2 $};
\hbox{$\scriptstyle 1\over2 $}(1-x)),
\end{equation}
where ${}_2F_1(a,b,c;z)$ is the hypergeometric function,
$C_n^\alpha(x)$ are the Gegenbauer polynomials and $\Gamma(x)$ is
the gamma function, we obtain
\begin{equation}
\fl \langle j,m;r|{\bi z}\rangle = e^{-\frac{1}{2}j(j+1)}\sqrt{2j+1}\,
\frac{(2|m|)!}{|m|!}\sqrt{\frac{(j-|m|)!}{(j+|m|)!}}\left(
\frac{-\varepsilon(m)z_1+{\rm i}z_2}{2}\right)^{|m|} C_{j-|m|}^{|m|+\frac{1}{2}}
(z_3),
\end{equation}
where $\varepsilon(m)$ is the sign of $m$.  Let us recall in the context of
the relations (5.14) that the polynomial dependence of the projection of
coherent states onto the discrete basis vectors, on the complex numbers
parametrizing those states is one of their most characteristic properties.
Clearly, the polynomials (5.14) should span via the ``resolution of
the identity operator'' the Fock-Bargmann representation.  We recall
that existence of such representation is one of the most important
properties of coherent states.  The problem of finding the
Fock-Bargmann representation in the discussed case of the coherent
states for a particle on a sphere is technically complicated and it will be
discussed in a separate work.  Finally, notice that the coherent states
$|{\bi z}\rangle$ are evidently stable under rotations.
\section{Coherent states and the classical phase space}
We now show that the introduced coherent states for a quantum
particle on a sphere are labelled by points of the classical phase
space, that is $T^*S^2$.  Referring back to eq.\ (4.16) and
taking into account the fact that the classical limit corresponds to
large $j$'s, we arrive at the following parametrization of ${\bi z}$
by points of the phase space:
\begin{equation}
{\bi z}=\cosh|{\bi l}|\,\frac{{\bi x}}{r}+{\rm i}\frac{\sinh|{\bi l}|}
{|{\bi l}|}\,{\bi l}\times \frac{{\bi x}}{r},
\end{equation}
where the vectors ${\bi l},\,{\bi x}\in{\Bbb R}^3$, fulfil
${\bi x}^2=r^2$ and ${\bi l}\bdot{\bi x}=0$, i.e.,
we assume that ${\bi l}$ is the classical angular momentum and ${\bi x}$ is the
radius vector of a particle on a sphere.  In accordance with the
formulae (4.15) and (4.13) the vector ${\bi z}$ satisfies ${\bi z}^2=1$.
Thus, the vector ${\bi z}$ is really parametrized by the points $({\bi x},{\bi l})$
of the classical phase space $T^*S^2$.

Consider now the expectation value of the angular momentum operator ${\bi J}$
in a coherent state.  The explicit formulae which can be derived
with the help of (3.7{\em a}), (3.8), (3.12) and (5.14) are too complicated
to reproduce them herein.  From computer simulations it follows that
\begin{equation}
\langle{\bi J}\rangle_{\bi z} =\frac{\langle {\bi z}|{\bi J}|{\bi z}\rangle}{\langle {\bi
z}|{\bi z}\rangle}\approx{\bi l}.
\end{equation}
Nevertheless, in opposition to the case with the circular motion, the
approximate relation (6.2) does not hold for practically arbitrary small
$|{\bi l}|$.  Namely, we have found that whenever $|{\bi l}|\sim1$, then
(6.2) is not valid.  Note that returning to dimension entities in
the formulae like (3.6) we measure $|{\bi l}|$ in the units of
$\hbar$, so in the physical units we deal rather with ${\bi L}=\hbar
{\bi l}$.  For $|{\bi l}|\ge10$ the relative error $|(\langle
J_i\rangle_{\bi z}-l_i)/\langle J_i\rangle_{\bi z}|$, $i=1,\,2,\,3$, is small.
More precisely, if $|{\bi l}|\sim10$, then $|(\langle
J_i\rangle_{\bi z}-l_i)/\langle J_i\rangle_{\bi z}|\sim$1 per cent.  In other
words, in the case of the motion on a sphere, the quantum fluctuations are not
negligible for $|{\bi L}|\sim$1 $\hbar$ and the description based on the
concept of the classical phase space is not adequate one.  However, it
must be borne in mind that the condition $|{\bi L}|\ge$ 10 $\hbar$, when (6.2)
holds is not the same as the classical limit $|{\bi l}|\to\infty$.  We only
point out that $10\,\hbar\approx 10^{-33}\,{\rm J}\cdot{\rm s}$.  It thus appears
that the parameter ${\bi l}$ in (6.2) can be identified with the classical angular
momentum divided by $\hbar$.

We now study the role of the parameter ${\bi x}$ in (6.1).  As with
the momentum operator ${\bi J}$ the explicit relations obtained by
means of (3.9) with $\zeta=0$, (3.12) and (5.14) are too
complicated to write them down herein.  The computer simulations
indicate that
\begin{equation}
\langle{\bi X}\rangle_{\bi z}=\frac{\langle{\bi z}|{\bi X}|{\bi z}\rangle}
{\langle {\bi z}|{\bi z}\rangle}\approx e^{-\frac{1}{4}}{\bi x}.
\end{equation}
It seems that the formal resemblance of the formula (6.3) and (2.19)
referring to the case with the circular motion is not accidental one.
The range of application of (6.3) is the same as for (6.2), i.e., $|{\bi
l}|\ge10$.  Because of the term $e^{-\frac{1}{4}}$, it appears that
the average value of ${\bi X}$ does not belong to the sphere with
radius $r$.  Proceeding analogously as in the case of the circle we introduce
the relative average value of ${\bi X}$ of the form
\begin{equation}
\langle\!\langle X_i\rangle\!\rangle_{\bi z}=\frac{\langle X_i\rangle_{\bi z}}
{\langle X_i\rangle_{{\bi w}_i}},\qquad i=1,\,2,\,3,
\end{equation}
where $|{\bi w}_i\rangle$ is a coherent state with
\begin{equation}
{\bi w}_k=\cosh|{\bi l}|{\bi n}_k+{\rm i}\frac{\sinh|{\bi l}|}{|{\bi l}|}
{\bi l}\times{\bi n}_k,\qquad k=1,\,2,\,3,
\end{equation}
where ${\bi n}_k$ is the unit vector along the $k$ coordinate axis
and ${\bi l}$ is the same as in (6.1).  In view of (6.3) and (6.4) we have
\begin{equation}
\langle\!\langle {\bi X}\rangle\!\rangle_{\bi z}\approx{\bi x}.
\end{equation}
Therefore, the relative expectation value $\langle\!\langle {\bi X}\rangle\!
\rangle_{\bi z}$ seems to be the most natural one to describe the average
position of a particle on a sphere.

We have thus shown that the parameter ${\bi x}$ can be immediately
related to the classical radius vector of a particle on a sphere.  As with the
case of the circular motion (see formulae (2.18) and (2.21)), we
interpret the relations (6.2) and (6.6) as the best possible
approximation of the classical phase space.  In this sense the
coherent states labelled by points of such deformed phase space are
closest to the classical ones.  The quantum fluctuations which are
the reason of the approximate nature of (6.2) and (6.6) are in our
opinion a characteristic feature of quantum mechanics on a sphere.

We finally remark that the discussion of the Heisenberg uncertainty
relations analogous to that referring to the circle (see section 2)
can be performed also in the case with the coherent states for a
particle on a sphere.  For example a counterpart of the
formula (2.22) is
\begin{equation}
(\Delta {\bi J})^2\ge\frac{1}{2}\frac{\frac{1}{2}{\rm Tr}\langle
V\rangle^2}{1-\frac{1}{2}{\rm Tr}\langle V\rangle^2},
\end{equation}
where according to eq.\ (4.2) we have $\langle V\rangle=\frac{1}{r}
\bsigma\bdot\langle {\bi X}\rangle$.  Such discussion as
well as the detailed analysis of the Heisenberg uncertainty relations
for the quantum mechanics on a compact manifold will be the subject of
a separate paper which is in preparation.
\section{Simple application: the rotator}
We now illustrate the actual treatment by the example of a free
twist 0 particle on a sphere, i.e.\ the rotator.  The corresponding
Hamiltonian is given by
\begin{equation}
\hat H=\hbox{$\scriptstyle 1\over2 $}{\bi J}^2.
\end{equation}
By (3.7{\em a}) the normalized solution of the Schr\"odinger equation
\begin{equation}
\hat H |E\rangle = E |E\rangle
\end{equation}
can be expressed by
\begin{equation}
|E\rangle= |j,m;r\rangle,\qquad E=\hbox{$\scriptstyle 1\over2 $}j(j+1).
\end{equation}
We now discuss the distribution of the energies in the coherent
state.  The computer simulations indicate that the function
\begin{equation}
p_{j,m}({\bi x},{\bi l})=\frac{|\langle j,m;r|{\bi z}\rangle|^2}{\langle
{\bi z}|{\bi z}\rangle},\qquad -j\le m\le j,
\end{equation}
determined by (5.14) and (6.1), which gives the probability
of finding the system in the state $|j,m;r\rangle$, when the system
is in the normalized coherent state $|{\bi z}\rangle/\sqrt{\langle{\bi
z}|{\bi z}\rangle}$, has the following properties.  For fixed
integer $m=l_3$ the function $p_{j,m}$ has a maximum at $j_{\rm max}$
coinciding with the integer nearest to the positive root of the equation
\begin{equation}
j(j+1)={\bi l}^2,
\end{equation}
(see Fig.\ 1).  Thus, it turns out that the parameter $\frac{1}{2}{\bi l}^2$
can be regarded as the energy of the particle.  Further, for fixed integer $j$ in $p_{j,m}({\bi
x},{\bi l})$ (see Fig.\ 2), such that (7.5) holds,
the function $p_{j,m}$ has a maximum at $m_{\rm max}$ coinciding
with the integer nearest to $l_3$.  It thus appears that the parameter $l_3$
can be identified with the projection of the momentum on the $x_3$ axis.
\section{Conclusion}
In this work we have introduced the coherent states for a quantum
particle on a sphere.  An advantage of the formalism used
is that the coherent states are labelled by points of the classical
phase space.  The authors have not found alternative constructions
of coherent states for a quantum mechanics on a sphere preserving
this fundamental property of coherent states.  As pointed out in
Sec.\ 6, the quantum fluctuations arising in the case of the motion
on a sphere are bigger than those taking place for the circular
motion.  This observation is consistent with the appearance of the
additional degree of freedom for the motion on a sphere.  We
remark that as with the particle on a circle, we deal within
the actual treatment with the deformation of the classical phase
space expressed by the approximate relations (6.2) and (6.6).  We
also point out that besides (6.2) and (6.6) the quasi-classical
character of the introduced coherent states is confirmed by the
behaviour of the distribution of the energies investigated in
section 7.  It seems that the approach introduced in this paper is not
restricted to the study of the quasi-classical aspects of the quantum
motion on a sphere.  For example, the results of this work would be of
importance in the theory of quantum chaos.  In fact, in this theory the kicked
rotator is one of the most popular model systems.  Because of the
well known difficulties in the analysis of the Heisenberg uncertainty relations
occuring in the case with observables having compact spectrum like
the position operator ${\bi X}$ satisfying the $e(3)$ algebra (3.4) we
have not studied them herein.  The analysis of the Heisenberg
uncertainty relations as well as the discussion of the case with a
nonvanishing twist will be performed in future work.

\Figures
\begin{figure}
\caption{The plot of $\ln p_{j,m}$ (see (7.5)), with fixed
$m=0$ and ${\bi z}$ given by (6.1), where ${\bi
x}=(0.412,0.412,0.812)$ and ${\bi l}=(8.124,-8.124,0)$.  Since
${\bi l}^2=132$, therefore $j_{\rm max}=11$ coincides with the
positive root of Eq.\ (7.5).}
\label{fig1}
\caption{The plot of $\ln p_{j,m}$ with ${\bi x}=(0.411,0.911,0.036)$
and ${\bi l}=(-17.490,7.490,10)$.  The fixed $j=21$ corresponds to the
positive root of (7.5), where ${\bi l}^2=462$.  The fuction has the
maximum at $m_{\rm max}=l_3=10$.}
\label{fig2}
\end{figure}
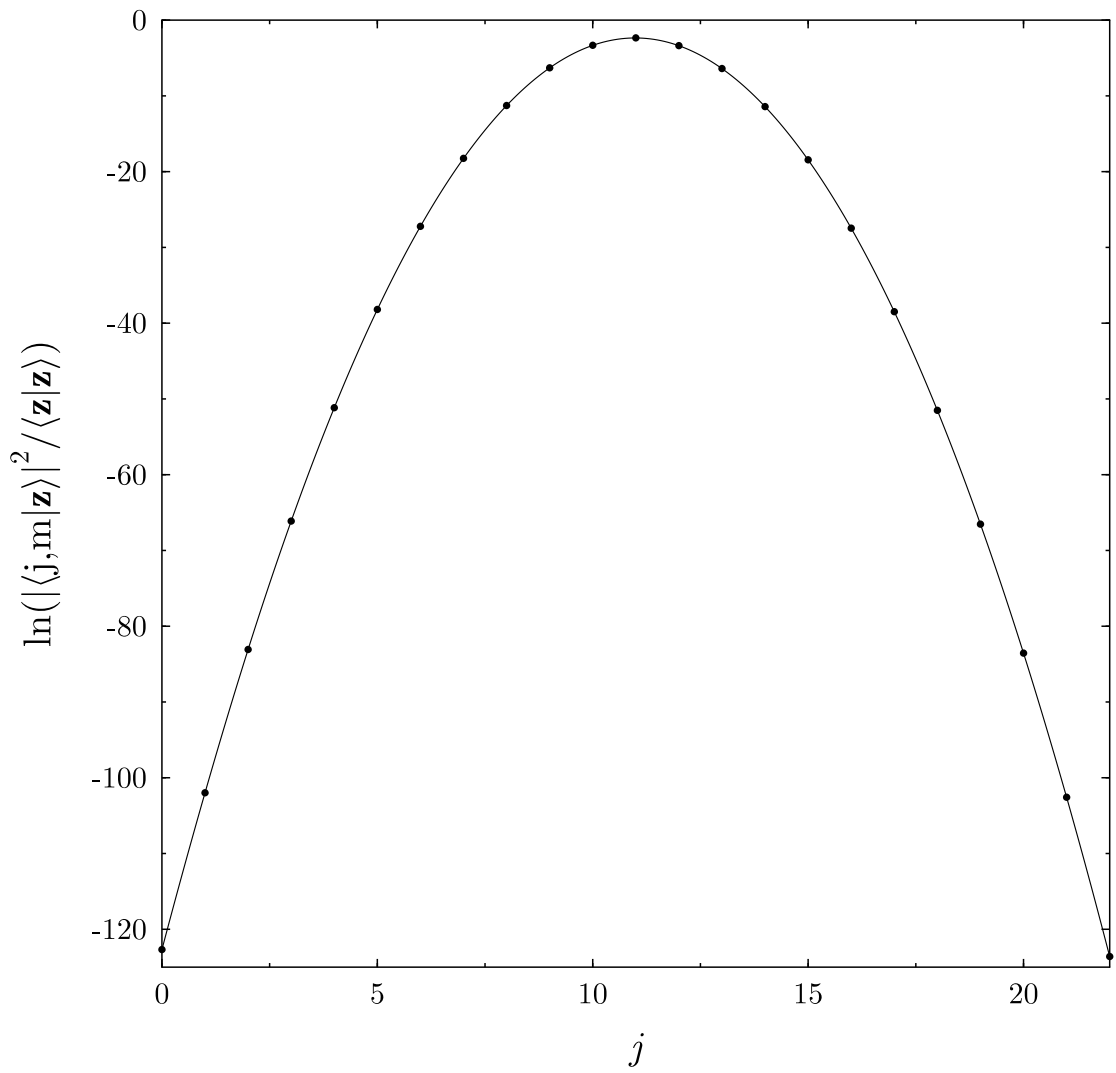
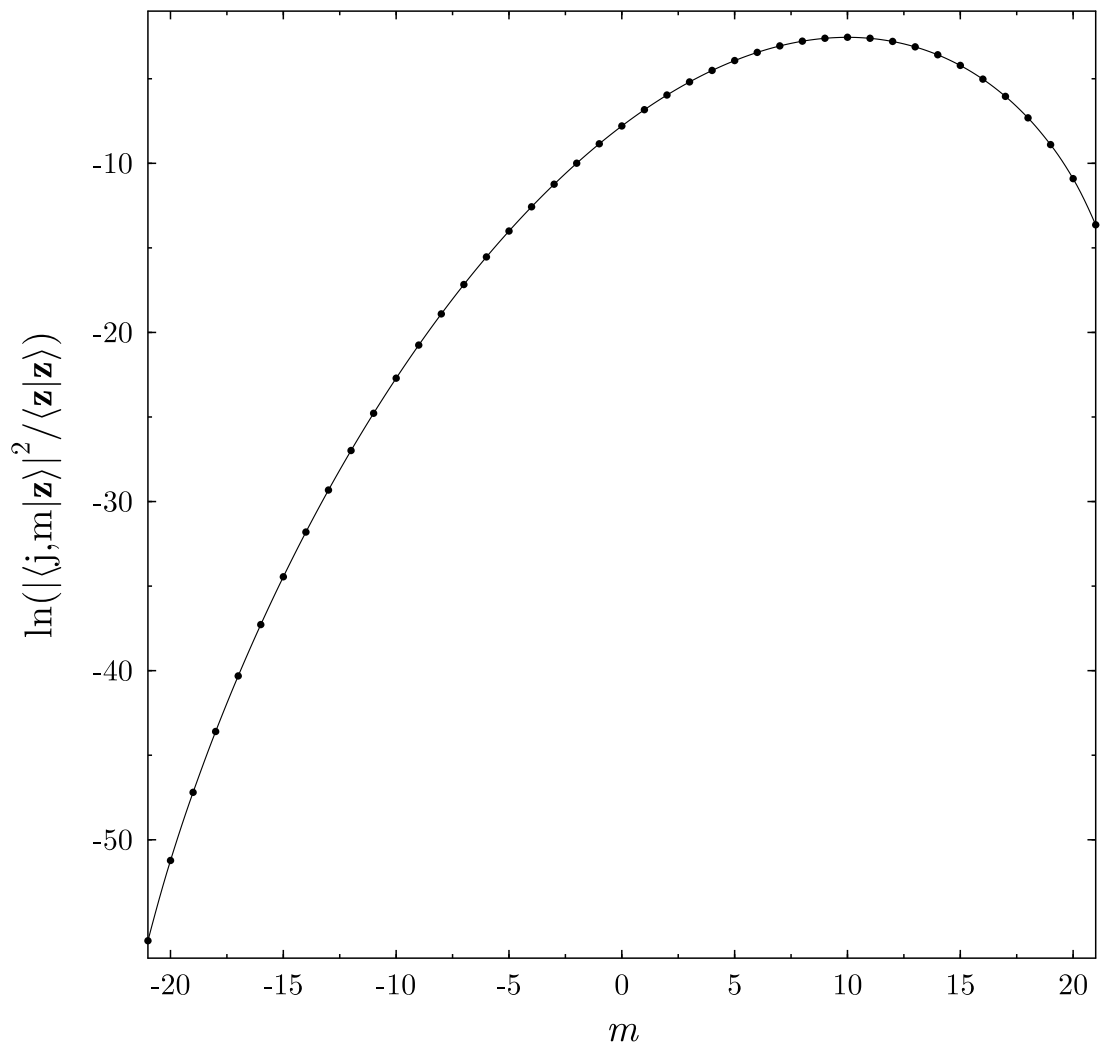
\end{document}